\iftrue
\documentclass[aps,prl,twocolumn, 
			   groupedaddress,superscriptaddress,
			   amsfonts,amssymb,amsmath,
			   citeautoscript,
			   a4paper]{revtex4-1}
\else
\documentclass[aps,pra,preprint,groupedaddress,superscriptaddress,notitlepage,
			   amsfonts,amssymb,amsmath,
			   citeautoscript,
			   a4paper]{revtex4-1}
\fi

\usepackage[pdftex]{hyperref}
\hypersetup{colorlinks,
			linkcolor={blue!75!black!80!yellow},
			citecolor={blue!75!black!80!yellow},
			urlcolor={blue!75!black!80!yellow},
			pdfstartview=FitH}
\usepackage{graphicx}
\usepackage{mathrsfs}
\usepackage{amsthm}
\usepackage{xspace}
\usepackage{braket}
\usepackage{xr}
\usepackage{gensymb} 
\usepackage{xcolor,soul}
\usepackage{stmaryrd} 
\usepackage[UKenglish]{babel}
\usepackage{placeins} 
\usepackage{siunitx}
\DeclareSIUnit[number-unit-product=]\percent{\char`\%} 
\usepackage{makecell}
\usepackage{diagbox}
\usepackage{float}
\usepackage{amsmath} 

\usepackage{txfonts}  
\usepackage{txfontsb} 
\usepackage{extarrows}
\usepackage{upgreek}

\newcommand{\iu}{\mathrm{i}}
\newcommand{\e}{\mathrm{e}}

\newcommand{\appropto}{\mathrel{\vcenter{
			\offinterlineskip\halign{\hfil$##$\cr
				\propto\cr\noalign{\kern2pt}\sim\cr\noalign{\kern-2pt}}}}}

\newcommand{\ie}{i.e.\@\xspace}  

\newcommand{\eg}{e.g.\@\xspace}

\newcommand{\kns}{\mathbf{k}\!-\!{\mathrm{NS}}}
\newcommand{\gx}{\gamma_x}
\newcommand{\gy}{\gamma_y}
\newcommand{\lx}{\lambda_x}
\newcommand{\ly}{\lambda_y}
\newcommand{\dxr}{d^x_{\mathrm{inter}}}
\newcommand{\dxa}{d^x_{\mathrm{intra}}}
\newcommand{\dyr}{d^y_{\mathrm{inter}}}
\newcommand{\dya}{d^y_{\mathrm{intra}}}

\makeatletter
\newcommand*{\addFileDependency}[1]{
  \typeout{(#1)}
  \@addtofilelist{#1}
  \IfFileExists{#1}{}{\typeout{No file #1.}}
}
\makeatother

\usepackage{scalerel}

\usepackage{textcomp} 
\usepackage{xifthen}
\usepackage{etoolbox}
\newboolean{togglecomments}
\newboolean{togglechanges} 

\setboolean{togglecomments}{true}
\setboolean{togglechanges}{false}

\newcommand{\comment}[2]{%
    \ifbool{togglecomments}%
    {\textcolor{blue!70!black}{\small\textsf{%
    \textsuperscript{\textsc{\textsf{\MakeLowercase{#1}}}}%
    [#2]}}} 
    {}}     
\newcommand{\swap}[2]{\ifbool{togglechanges}
    {#2}  
    {\textcolor{red!70!black}{[#1]}\textrightarrow{}\textcolor{green!50!black}{[#2]}}}
\newcommand{\remove}[1]{\ifbool{togglechanges}
    {}    
    {\textcolor{red!70!black}{#1}}}
\newcommand{\inset}[1]{\ifbool{togglechanges}
    {#1}  
    {\textcolor{green!50!black}{#1}}}
\newcommand{\optional}[1]{\ifbool{togglechanges}
    {#1}  
    {\textcolor{yellow!50!orange!80!gray}{#1}}}
\newcommand{\citeremind}[1]{%
    [\textcolor{blue!75!black!80!yellow}{
        $\blacksquare$%
           \ifthenelse{\isempty{#1}}
               {}
               {\textsuperscript{\textsf{#1}}}%
        }]\xspace}
\newcommand{\todo}[1]{
    \textcolor{orange!80!yellow!95!black}{\textbf{[}%
        \ifthenelse{\isempty{#1}}%
        {\text{$\blacksquare$}}%
        {{\small\textsf{#1}}}%
        \textbf{]}}}
        
\newcommand{\hkuaffil}{\footnotesize Department of Physics and HK Institute of Quantum Science and Technology, The University of Hong Kong, Pokfulam, Hong Kong, China}
\newcommand{\usstaffil}{\footnotesize College of Optical-Electrical Information and Computer Engineering,
University of Shanghai for Science and Technology, Shanghai 200093, China}
\newcommand{\ssmetaffil}{\footnotesize Shanghai Szon Mechanical $\&$ Electrical Technology Co., LTD, Shanghai, China}

\begin{document}

\title{Acoustic higher-order topological insulator from momentum-space nonsymmorphic symmetries}

\author{Jinbing Hu}
\email{hujinbing@usst.edu.cn}
\affiliation{\usstaffil}
\affiliation{\hkuaffil}
\author{Kai Zhou}
\affiliation{\usstaffil}
\author{Tianle Song}
\affiliation{\usstaffil}
\author{Xuntao Jiang}
\affiliation{\ssmetaffil}
\author{Songlin Zhuang}
\affiliation{\usstaffil}
\author{Yi~Yang}
\email{yiyg@hku.hk}
\affiliation{\hkuaffil}

\begin{abstract}
Momentum-space nonsymmorphic symmetries, stemming from the projective algebra of synthetic gauge fields, can modify the manifold of the Brillouin zone and lead to a variety of topological phenomena. 
We present an acoustic realization of higher-order topological insulators (HOTIs) protected by a pair of anticommutative momentum-space glide reflections.
We confirm the presence of momentum-space glide reflection from the measured momentum half translation of edge bands and their momentum-resolved probability distribution using a cylinder geometry made of acoustic resonator arrays.
In particular, we observe both intrinsic and extrinsic HOTI features in such a cylinder: hopping strength variation along the open boundary leads to a bulk gap closure, while that along the closed boundary results in an edge gap closure.
In addition, we confirm the presence of quadrupole corner modes with transmission and field distribution measurements.
Our observation enriches the study of topological physics of momentum-space nonsymmorphic symmetries.

\end{abstract}

\maketitle
Porjective symmetries have recently been shown to give rise to intriguing topological phenomena~\cite{zhang2023general,chen2023classification,zhao2021switching,xue2023stiefel,meng2023spinful,herzog2023hofstadter,liu2023mobius,shiozaki2015z,zhao2020z,xue2022projectively,li2022acoustic,long2024non,jiang2023photonic}.
The presence of synthetic gauge fields projectively modifies the form of conventional symmetries, such as translation and reflection symmetries, and their associated symmetry algebra, which can protect  M\"{o}bius edge modes~\cite{zhao2020z,xue2022projectively,li2022acoustic} and Klein-bottle topological insulators~\cite{chen2022brillouin,tao2024higher,zhu2024brillouin}.
One particular kind of projective symmetry is the momentum-space nonsymmorphic ($\kns$) reflection, which besides flipping the sign of a momentum, also performs a half translation along the reflection plane~\cite{chen2022brillouin,tao2024higher,li2023acoustic}. 
This momentum-space glide reflection operation is complementary but different from its real-space counterparts that have been extensively studied for stabilizing topological insulators and semimetals~\cite{lu2016symmetry,wang2016hourglass,wang2020symmetry,cheng2020discovering,zhang2020symmetry,liu2022geometric}.
The addition of half translation in $\kns$ reflection has several consequences.
First, the half translation removes the reflection axis and transforms the conventional reflection into a free operator that can modify the manifold of Brillouin zone (BZ) from a torus into a Klein bottle~\cite{chen2022brillouin}.
Second, due to the non-orientability of Brillouin zone, $\kns$ reflection can also protect topological Weyl semimetals that obey a no-go theorem~\cite{fonseca2024weyl} that is distinct from its counterpart for Weyl semimetals on the Brillouin torus~\cite{lv2015experimental,soluyanov2015type,armitage2018weyl,wang2021realization}.
Third, $\kns$ reflection constrains bulk polarization and Wannier-sector polarization nonlocally, and a pair of anti-commutative $\kns$ reflection can further stabilize intrinsic higher-order topological insulators (HOTIs) on a real projective plane~\cite{hu2024higher}.

By generalizing the electric dipole moment~\cite{resta1992theory,king1993theory,vanderbilt2018berry} to higher electric multipole moments, higher-order topological insulators (HOTIs) were recently defined~\cite{benalcazar2017quantized,benalcazar2017electric,schindler2018higher} and experimentally demonstrated in various physical platforms (\eg Refs.~\cite{peterson2018quantized,mittal2019photonic,qi2020acoustic,ni2020demonstration,xue2020observation,serra2018observation,imhof2018topolectrical}). 
According to whether the phase transition is related to bulk gap closure or not, HOTIs are ramified into intrinsic and extrinsic ones~\cite{khalaf2021boundary,geier2018second,queiroz2019partial} that are respectively featured with symmetry-protected topological phase (SPTP) and boundary-obstructed topological phase (BOTP)~\cite{khalaf2021boundary,du2022acoustic,chen2022experimental}.
In previous HOTI studies, SPTP and BOTP are typically considered mutually exclusive to each other once a model is constructed; nevertheless, it was recently predicted that in a HOTI protected by $\kns$ reflections, BOTP could appear \emph{within} the interior of an SPTP~\cite{hu2024higher}.
Despite that higher-order topological insulators stabilized by projective translation and reflection symmetries have been observed~\cite{tao2024higher,lai2024real},
the occurrence of boundary obstruction within an SPTP, together with the associated characterization of $\kns$ glide reflection from boundary features, has not been experimentally demonstrated. 

In this work, we realize a HOTI protected by a pair of anticommutative momentum-space nonsymmorphic symmetries. Based on acoustic resonator array structures, We confirm the realization of $\kns$ glide reflection from the momentum-resolved probability density of edge bands on a cylinder geometry, demonstrate the appearance of boundary obstruction within the interior of an SPTP, and observe the associated corner modes stabilized by $\kns$ glide reflections.
Acoustic resonator array structures are fabricated into two geometries: a planar full open structure for observing bulk, edge, and corner modes and a cylinder, which preserves the $\kns$ reflection along its periodic direction.
The measured edge bands reveal a momentum half translation between the edge modes localized at the opposite ends of the sample, which is also verified by their momentum-resolved probability density.
Furthermore, we observe both BOTP and SPTP phase transitions, respectively, by changing the intracell coupling strength along either the closed or open boundaries of the cylinder structure.

\begin{figure*}[hbtp]
    \centering
    \includegraphics[width=\linewidth]{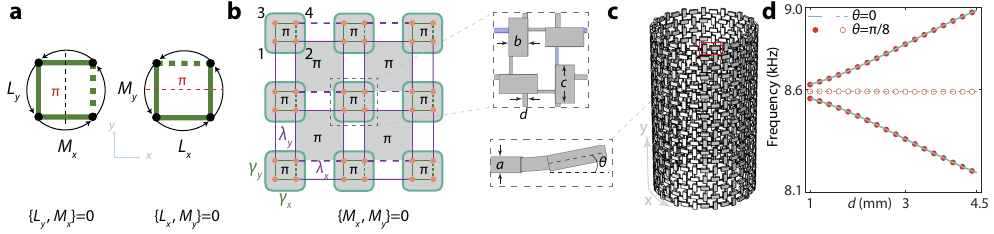}
    \caption{\textbf{Lattice model featuring nonsymmorphic reflections in momentum space and its acoustic realization.}     
    \textbf{a.} Threading $\pi$ flux projectively modifies the conventional reflections $m_i~(i=x,y)$ into $\kns$ ones $M_i~(i=x,y)$, which anticommute with the translation symmetry $L_j~(j=y,x)$ along the other direction. 
    \textbf{b.} Model with checkerboard $\pi$ flux enables anti-commutative $M_x$ and $M_y$. Solid and dashed lines indicate real positive and negative hoppings, respectively. The inset displays the implementation of an acoustic lattice unit cell, where blue (gray) tubes indicate negative (positive) hoppings.
    \textbf{c.} A cylinder structure of the lattice model preserves the translation symmetry in $x$ direction, thus, in turn, obeys $\kns$ $M_y$. The inset shows the bending design of the resonator connection along the $x$ direction, where $\theta$ is the bending angle.
    \textbf{d.} The splitting frequencies of two perpendicularly arranged identical resonators as a function of tube width $d$ for $\theta=\SI{0}{\degree}$ (green lines) and $\theta=\SI{11.25}{\degree}$ (red solid circles), respectively.
    In the middle, the near horizontal dashed line and hollow circles are the averages of the split frequencies, which indicate that the on-site potentials of resonators in both curved and planar connections are uniform with little variation under tube width changes.    
    }
\label{fig:lattice_model_curved_connection}
\end{figure*}

\emph{Model construction.}
The basic building blocks of the system are shown in Fig.~\ref{fig:lattice_model_curved_connection}a.
Due to the presence of $\pi$ gauge flux, the conventional reflection $m_x$ ($m_y$) is projectively modified as $M_x=U\mathcal{L}_{\mathbf{G_y}/2}m_x$ ($M_y=V\mathcal{L}_{\mathbf{G_x}/2}m_y$), where $U~(V)$ is an unitary matrix, $\mathcal{L}_{\mathbf{G_i}/2} (i=x,y)$ denotes the half translation along $i$ axis~\cite{chen2022brillouin}.
We construct the square lattice model with a checkerboard $\pi$ flux pattern (Fig.~\ref{fig:lattice_model_curved_connection}b), which is featured with both $\kns$ $M_x$ and $M_y$~\cite{hu2024higher}. The associated Bloch Hamiltonian is
\begin{align}\label{eq:bloch_H}
    H(k_x,k_y)=\left[ \begin{array}{cccc}    
                     0 & a_+ & b_+ & 0\\
                     a_+^* & 0 & 0 & b_-\\
                     b_+^* & 0 & 0 & a_-\\
                     0 & b_-^* & a_-^* & 0\\
                     \end{array}
              \right],
\end{align}
with $a_{\pm}=\gx\pm \lx \e^{-\iu k_x}$, $b_{\pm}=\pm \gy+\ly \e^{-\iu k_y}$, and $*$ denoting complex conjugate. Here, $\gx$ and $\gy$ ($\lx$ and $\ly$) represent the intra-cell (inter-cell) hopping amplitudes along $x$ and $y$, respectively. Without loss of generality, we set unity lattice constants and $\lx=\ly=1$. 
For the chosen coupling arrangement in Fig.~\ref{fig:lattice_model_curved_connection}b, the $\kns$ $M_x$ and $M_y$ have the form: $M_x=\sigma_3\otimes\tau_1$, $M_y=\sigma_1\otimes\tau_0$, where $\sigma$'s and $\tau$'s are Pauli matrices acting on sites along $y$ and $x$, respectively. 
The gauge flux pattern in Fig.~\ref{fig:lattice_model_curved_connection}b enables the anti-commutativity of $M_x$ and $M_y$, which together stabilize an intrinsic higher-order topological phase manifested by a nontrivial quadrupole moment, corner modes, and bulk gap closing across the phase boundary~\cite{hu2024higher}.
Aside from Fig.~\ref{fig:lattice_model_curved_connection}b, there exists a complementary way to insert a checkerboard $\pi$ flux pattern in the square lattice~\cite{wang2023chessboard,tao2023quadrupole}, which also generates a pair of $\kns$ $M_x$ and $M_y$; however, in that case, the $\kns$ $M_x$ and $M_y$ are commutative and the quadrupole moment is trivial (see Sec.~S1 of \cite{SM_note}).

\emph{Sample design.}
In the experiment, the checkerboard $\pi$ flux pattern in Fig.~\ref{fig:lattice_model_curved_connection}b is implemented via coupled acoustic resonator structures~\cite{xue2020observation,xue2022projectively,xue2023stiefel,hu2024higher}, where the resonators emulating the lattice site are designed to work at the dipolar mode near frequency $\SI{8587}{Hz}$ (see Sec.~S2 of \cite{SM_note}). 
By perpendicularly arranging two identical resonators, the hopping phase of these two identical resonators is controllable through left or right localizing the connecting tube~\cite{xue2020observation,xue2022projectively,xue2023stiefel}. The coupling strength can be tuned via the width of the cross-section of the tube and its position with respect to the center of the resonator.

\begin{figure}
    \centering
    \includegraphics[width=1\linewidth]{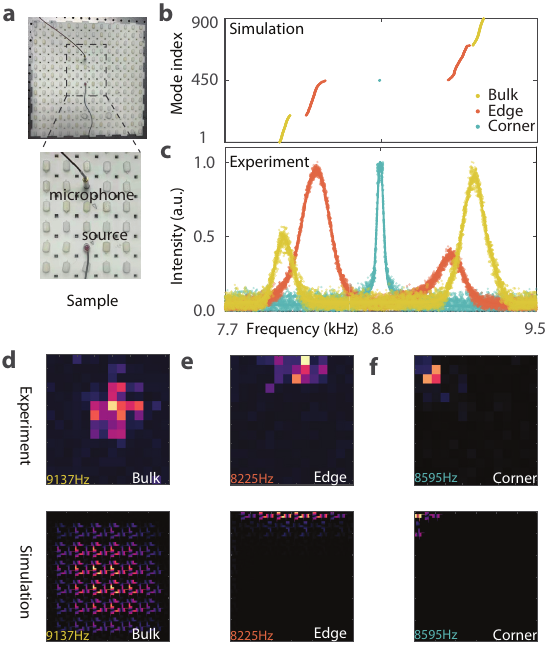}
    \caption{
    \textbf{Corner mode measurement.}
    \textbf{a.} Setup for measuring the spectral response. The inset is a zoom-in for better showing the source-detector positioning of the sound source and the probing microphone.
    \textbf{b-c.} Simulated (b) and measured (c) spectra of bulk (yellow), edge (brown), and corner (green) modes.
    \textbf{d-f.} Measured (upper row) and simulated (bottom row) pressure intensity distributions for bulk, edge, and corner modes with frequencies being $\SI{9137}{Hz}$, $\SI{8225}{Hz}$, and $\SI{8595}{Hz}$, respectively.
    }  
    \label{fig:measured_bulk_edge_corner_spectrum}
\end{figure}

In addition to the full-open planar structures (Fig.~\ref{fig:measured_bulk_edge_corner_spectrum}) for corner mode characterizations, here we also implement cylinder structures (open and closed on two directions, respectively) to experimentally validate the momentum half translation of the $\kns$ reflection.
For this aim, two types of connections among resonators should be designed as equivalent: along the open $y$ direction, there is a planar connection where the two neighboring resonators and the connecting tube are co-planar, and along the closed $x$ direction, there is a curved connection, where the two neighboring resonators are connected via a curved tube described by a parametric angle $\theta$ (see inset of Fig.~\ref{fig:lattice_model_curved_connection}c).
The design challenge is that despite linked with distinct planar and curved connections, the resonators should maintain the periodicity and the same pressure distribution along both directions (see Sec.~S2 of \cite{SM_note}), such that the fabricated cylinder can faithfully realize the Hamiltonian [Eq.~\eqref{eq:bloch_H}] in its cylinder geometry.
To this end, we optimize the geometry of the resonators and their connections and arrive at the set of parameters (Fig.~\ref{fig:lattice_model_curved_connection}b): $a=\SI{5.2}{\mm}$, $b=\SI{10}{\mm}$, $c=\SI{20}{\mm}$, and the connections with $\theta=0$ and $\theta=\pi/8$ are equivalent for tube width $d$ in the range of \SIrange{1}{4.5}{\mm} when the tube length is fixed at $\SI{9.3}{mm}$, as shown in Fig.~\ref{fig:lattice_model_curved_connection}d. 
Unless noted otherwise, the simulation model and realistic sample are fabricated based on the optimization above.
Fig.~\ref{fig:lattice_model_curved_connection}d also shows that the central frequency of resonators is designed uniform under various tube length such that we can unify the on-site potential of resonators. 

\begin{figure}
    \centering
    \includegraphics[width=1\linewidth]{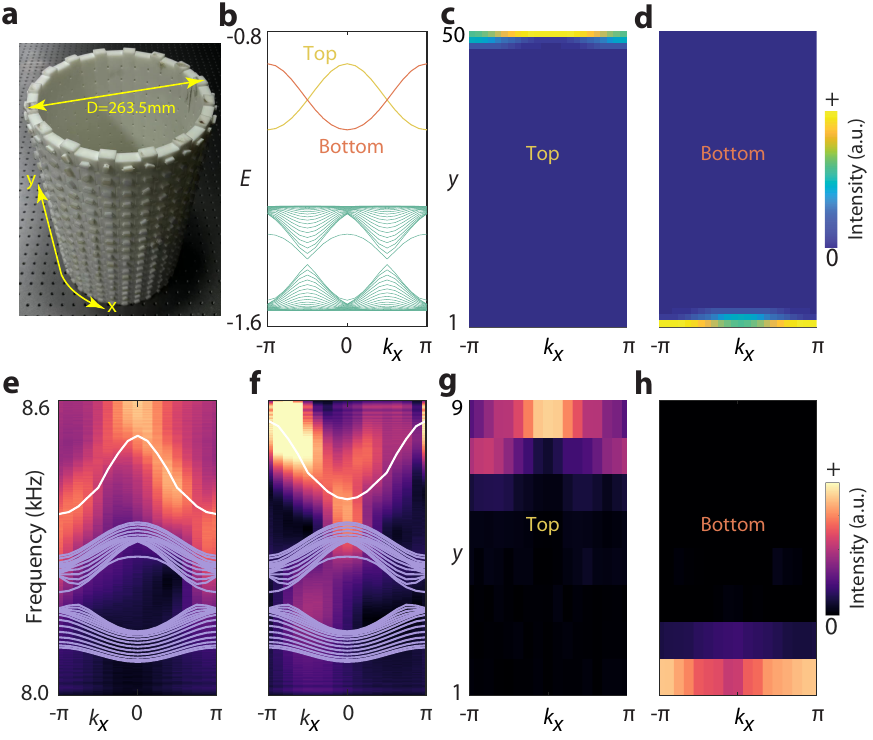}
    \caption{
    \textbf{Measured edge bands and their momentum-resolved probability density on an acoustic cylinder.}
    \textbf{a.} Photo of the cylinder with the size of $16\times9$ unit cells along $x$ and $y$ respectively. The connecting tube widths of the cylinder along $x$ and $y$ axes are: $\dxa=\SI{1.7}{mm}$, $\dxr=\SI{3.0}{mm}$, $\dya=\SI{1.7}{mm}$, $\dyr=\SI{4.5}{mm}$. 
    \textbf{b.} Projected bands of the $y$-open cylinder with hopping parameters $\gx=\gy=0.1$, $\lx=\ly=1.0$.
    \textbf{c-d.} The momentum-resolved probability density of edge bands in b. 25 unit cells are assumed along the $y$ axis. 
    \textbf{e-f.} Measured edge bands of the acoustic cylinder under different source conditions. The source is positioned at the top (e) and bottom (f) edge of the cylinder, repsectively. The purple and white curves are the theoretical bulk and edge bands from acoustic full-wave simulations. 
    \textbf{g-h.} Measured momentum-resolved probability density of the edge bands (corresponding to white curves in e and f, respectively). 
    }  
    \label{fig:measured_half_translation}
\end{figure}

\begin{figure}
    \centering
    \includegraphics[width=1\linewidth]{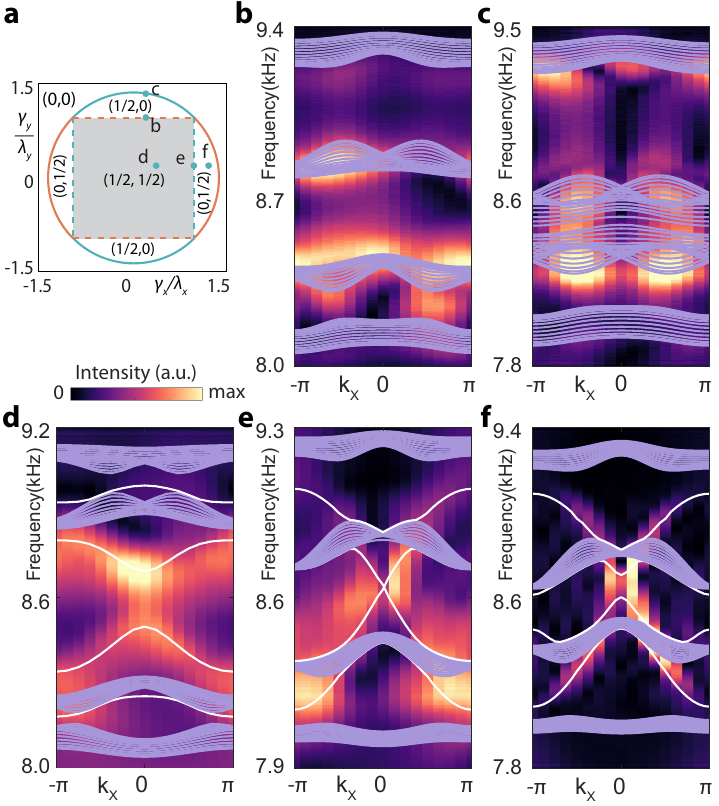}
    \caption{\textbf{BOTP and SPTP transitions along the closed and open boundaries.}
    \textbf{a.} Phase diagram of edge polarization $(p_x^e,p_y^e)$. Red and green lines correspond to $x$-open and $y$-open cylinders, respectively. Solid and dashed lines refer to bulk and edge gap closure, respectively.
    \textbf{b-f.} Measured band structure of the $y$-open acoustic cylinder with structure parameters indicated by their associated points in a. The purple and white curves are the theoretical bulk and edge bands from acoustic full-wave simulations.
    The bulk gap closure from b to c indicates the SPTP phase transition of edge polarization along the open boundary.
    The edge band gap closes and reopens in d, e, and f, indicating the BOTP phase transition of edge polarization along the closed boundary.
    The structural parameters for the $y$-open acoustic cylinder are $\dxa=(1.7, 1.7, 1.7, 3.0, 3.5)~\SI{}{mm}$, $\dxr=(4.5, 4.5, 3.0, 3.0, 3.0)~\SI{}{mm}$, $\dya=(3.0, 4.0, 1.7, 1.7, 1.7)~\SI{}{mm}$, $\dyr=(3.0, 3.0, 4.5, 4.5, 4.5)~\SI{}{mm}$ for b-f, respectively.
    }  
    \label{fig:measured_botp_sptp}
\end{figure}

\emph{Half translation of $\kns$ reflection.}
We validate the $\kns$ reflection symmetry using a $y$-open cylinder (Fig.~\ref{fig:measured_half_translation}a). 
In this cylinder, because the translation symmetry along $x$ remains intact, $\kns$ $M_y$ is preserved and pairs both bulk and edge bands with their half-translation partners.
Detached from the bulk bands (see Fig.~\ref{fig:measured_half_translation}b), the top-localized and bottom-localized edge bands (red and yellow curves in Fig.~\ref{fig:measured_half_translation}b) have the same dispersion relation but differ by a half translation, which serves as a convenient observable to confirm the existence of $\kns$ reflection.
This half translation of $\kns$ $M_y$ is also manifested through the predicted momentum-resolved probability density of edge bands, as shown by the comparison of Figs.~\ref{fig:measured_half_translation}c and d.

Experimentally, an acoustic cylinder with the size of $16\times 9$ unit cells along $x$ and $y$ direction, respectively, is fabricated (see Fig.~\ref{fig:measured_half_translation}a).
To measure the top-localized and bottom-localized edge bands, we excite the cylinder with a broadband sound source localized at the top or bottom resonator, and simultaneously record the time-domain signal from the same resonators at each cell.
By Fourier transform on the recorded time-space data (see Sec.~S3 of \cite{SM_note}), the dispersion of the measured top and bottom edge bands are obtained, as shown by the hot maps in Figs.~\ref{fig:measured_half_translation}e and f, where the discrete curves are simulation results.
The measured edge bands agree well with the theoretical ones (white lines) in the simulation, both of which exhibit momentum half translation between the top and bottom edge bands.
The inhomogeneous distribution of Fourier intensity in Fig.~\ref{fig:measured_half_translation}f may come from the fabrication disorders (see Sec.~S4 of \cite{SM_note}).
In addition, the momentum-resolved probability density of the top and bottom edge bands are also measured by scanning the detector (position fixed at the same resonator with each unit cell) across the whole cylinder, and the results are plotted in Figs.~\ref{fig:measured_half_translation}g and h (see Sec.~S3 of \cite{SM_note}), which also verify the half translation between these two edge bands. 

\emph{BOTP inside SPTP.}
As predicted in Ref.~\cite{hu2024higher}, boundary obstruction can appear within the interior of SPTP for HOTIs stabilized by $\kns$ symmetries.
In model Fig.~\ref{fig:lattice_model_curved_connection}b, BOTP can be introduced inside the intrinsic quadrupole phase by opening one pair of boundaries.
Fig.~\ref{fig:measured_botp_sptp}a shows the phase diagram of edge polarization~\cite{benalcazar2017electric,benalcazar2017quantized} (see Sec.~S5 of~\cite{SM_note} for computation) of model Fig.~\ref{fig:lattice_model_curved_connection}b in cylinder geometry, where the phase transition can occur via either bulk or edge gap closures: hopping strength variation along the \emph{open} direction leads to a \emph{bulk} gap closure (Fig.~\ref{fig:measured_botp_sptp}b-c that cross a solid arc in Fig.~\ref{fig:measured_botp_sptp}a), while that along the \emph{periodic} direction results in an \emph{edge} gap closure (see Fig.~\ref{fig:measured_botp_sptp}d-f that cross a dashed line in Fig.~\ref{fig:measured_botp_sptp}a).

We demonstrate such BOTP and SPTP transitions by measuring the band structures under various coupling strengths along the closed and open directions of the fabricated acoustic cylinders, respectively. 
In Fig.~\ref{fig:measured_botp_sptp}b-f, the measured results are plotted overlaid with theoretical bands (discrete curves; edge modes in white and bulk modes in purple) of $y$-open cylinder structures.
When $\dya=\dyr$ the bulk bands exhibit a large gap (Fig.~\ref{fig:measured_botp_sptp}b), which closes as we increase $\dya$ (Fig.~\ref{fig:measured_botp_sptp}c), indicating the occurrence of SPTP transition.
In contrast, the scenario along closed boundary (\ie, $x$ axis) is different. When $\dxa<\dxr$, a large edge band gap is observed, (Fig.~\ref{fig:measured_botp_sptp}d), which closes when $\dxa=\dxr$ (Fig.~\ref{fig:measured_botp_sptp}e) and reopens once $\dxa$ exceeds $\dxr$ (Fig.~\ref{fig:measured_botp_sptp}f).
Jointly, Figs.~\ref{fig:measured_botp_sptp}d-f display the process of BOTP transition.

\emph{Measurement of corner modes.}
The bulk of the model in Fig.~\ref{fig:lattice_model_curved_connection}b exhibits a quadrupole insulating phase~\cite{he2020quadrupole,zhang2020symmetry,zhou2020twisted,lin2020anomalous,zhou2024realization}, which is manifested by gapped corner modes under full open boundary condition when $|\gx|<|\lx|$ and $|\gy|<|\ly|$~\cite{hu2024higher}. 
Based on the building block in Fig.~\ref{fig:lattice_model_curved_connection}, we fabricate the planar acoustic sample with a size of $7\times7$ unit cells using three-dimensional printing techniques.
For the scanning of the sound field, each resonator is perforated with a hole (which is sealed when not used) to insert the sound source or microphone.
The experiment setup for spectra measurement is shown in Fig.~\ref{fig:measured_bulk_edge_corner_spectrum}a, where the acoustic source and microphone are localized in the bulk of the sample with two resonators in between; the source and microphone will be located at the edge or corner of the sample when the edge or corner spectra are measured. The details of the source-detector configurations for the bulk, edge, and corner spectra can be found in S6 of \cite{SM_note}.
As shown in Fig.~\ref{fig:measured_bulk_edge_corner_spectrum}c, the bulk (yellow dots) and edge (red dots) spectra feature a common wide gap at \SIrange{8400}{8860}{\Hz}, at the center of which a corner (green dots) spectral peak (around $\SI{8596}{Hz}$) appears.
The corresponding simulation results are plotted in Fig.~\ref{fig:measured_bulk_edge_corner_spectrum}b, in good agreement with the measured spectra.
Moreover, the field distributions at the specific peak frequencies of bulk, edge, and corner spectra, \ie, $\SI{9137}{Hz}$, $\SI{8225}{Hz}$, $\SI{8595}{Hz}$, respectively, are measured by field scanning across all resonators. 
The results are displayed in Figs.~\ref{fig:measured_bulk_edge_corner_spectrum}d-f, and 
are in good agreement with the associated numerical simulations (bottom row in Fig.~\ref{fig:measured_bulk_edge_corner_spectrum}d-f) under the same excitation frequencies (see S6 of \cite{SM_note} for the measurement details of the pressure distribution).

\emph{Conclusions.}
In summary, we experimentally realize a pair of momentum-space glide reflections that protect a higher-order topological insulating phase in an acoustic resonator array. 
We observe the momentum half translation of the symmetry, the boundary obstruction within a symmetry-protected phase, and the resulting corner modes.
The findings here could be relevant for a wider range of momentum-space nonsymmorphic symmetries that can be flexibly synthesized from inhomogeneous gauge fields in both Hermitian and non-Hermitian systems.

\begin{acknowledgments}
The authors thank Hengbin Cheng, Andr\'e Grossi Fonseca, Yixin Sha, Sachin Vaidya, Yong Xu, Haoran Xue, and Mou Yan for fruitful discussions and \mbox{Yangyang} Zhao for experimental help. 
\end{acknowledgments}

%

\end{document}